\documentclass[draftcls,12pt,onecolumn]{IEEEtran}
\usepackage{geometry}
\usepackage{epsfig}
\usepackage{amsmath,amssymb}
\usepackage{xcolor}
\usepackage{algorithm}
\usepackage{algorithmic}
\usepackage{multirow}
\usepackage{graphicx}
\usepackage{stfloats}
\usepackage{amsfonts}
\usepackage{mathrsfs}
\usepackage[noblocks]{authblk}
\usepackage[compress]{cite}
\usepackage{bm}
\usepackage{rotating}
\usepackage{changebar}
\usepackage{longtable}
\usepackage{chngpage}
\usepackage{array}
\usepackage{booktabs}
\usepackage[printonlyused]{acronym}
\begin{document}

\title{\LARGE  Resource Allocations for Secure Cognitive Satellite Terrestrial Networks
\author{Bin Li, Zesong Fei,~\IEEEmembership{Senior Member,~IEEE}, Xiaoming Xu, and Zheng Chu,~\IEEEmembership{Member,~IEEE}}
\thanks{B. Li and Z. Fei are with the School of Information and Electronics, Beijing Institute of Technology, Beijing 100081, China (email: feizesong@bit.edu.cn).}
\thanks{X. Xu is with the National Digital Switching System Engineering \& Technological Research Center, Zhengzhou 450002, China (email: ee\_xiaomingxu@sina.com).}
\thanks{Z. Chu is with the School of Science and Technology, Middlesex University, London, NW4 4BT, U.K. (e-mail: z.chu@mdx.ac.uk).}
}

\maketitle

\begin{abstract}
Cognitive satellite-terrestrial networks (STNs) have been recognized as the promising architecture for addressing spectrum scarcity in the next generation communication networks.
In this letter, we investigate the secure transmission schemes in the cognitive STNs where the interference from terrestrial base station (BS) is introduced to enhance security of the satellite link.
Our objectives are to minimize the transmit power by jointly optimizing the cooperative beamforming and artificial noise (AN) while guaranteeing the secrecy rate constraint for satellite link, the information rate constraint for terrestrial link and the total transmit power constraint. Both scenarios of the perfect and imperfect channel cases are respectively considered.
These constraints make the optimization problems formulated non-convex and challenging,
which are efficiently solved via certain transformations to be formulated more tractable versions, respectively.
Numerical results are provided to corroborate the proposed schemes.
\end{abstract}


\section{Introduction \label{a}}
Since the communication performance of the satellite networks are impacted by the masking effects between the satellite (SAT) and terrestrial nodes, the emerging cognitive satellite-terrestrial networks (STNs) are expected to bring many exciting applications, ranging from the utilization of scarce satellite spectrum resource to security improvements \cite{An2016JSAC}.
The promising network architecture allows the satellite network and the terrestrial network operating at the same frequency band, which offers new possibilities and challenges\cite{Li2016CL}.

Recently, the concept of cognitive STNs as a new paradigm has been investigated \cite{Vassaki2013CL,Lagunas2015TCCN,Shi2017CL}. To be specific,
\cite{Vassaki2013CL} first studied a resource management scheme in the novel architecture, while the cognitive exploitation
mechanisms were presented in \cite{Lagunas2015TCCN} for both the forward link and return link scenarios.
Considering the real-time applications in the cognitive STNs, \cite{Shi2017CL} discussed two power control schemes for maximizing the delay-limited capacity
and outage capacity of satellite user.

Owing to the broadcasting nature and inherent openness of the satellite networks, any receiver situated within the coverage of the SAT can naturally receive the transmitted signal.
As a result, the security issues are of paramount importance in the satellite communications, especially in military applications \cite{Lei2011TIFS,Zheng2012TWC}.
In recent years, physical-layer security (PLS), which is based on the physical-layer
characteristics of the wireless channels, has been proposed to realize secure communications.
The concept of PLS was firstly applied to satellite communications by Lei \textit{et al.}\cite{Lei2011TIFS}. Then, Zheng \textit{et al.} \cite{Zheng2012TWC} designed an optimal transmit beamforming scheme to realize secure transmission of multibeam satellite communications and Yan \textit{et al.} \cite{Yan2016WOCC} proposed a secure transmit design in
the satellite-terrestrial relay networks. On the other hand, the PLS in the cognitive STNs have also been attracted substantial attentions. Specially,
Yuan \textit{et al.} \cite{Yuan2016VTC} designed the optimal beamforming jointly in the cognitive STNs for maximizing the secrecy rate of primary user (PU), while the secrecy performance of
PU was analyzed in \cite{An2016JSAC} under two realistic scenarios, i.e., the CSI of eavesdropper was available or unavailable at SAT.
However, applications of PLS in cognitive STNs are relatively limited, and this motivates our current works.

In this letter, we focus on the secure cooperative transmission scheme for the cognitive STNs, where the satellite network acts as primary network and the terrestrial network acts as secondary network.
Unlike \cite{Zheng2012TWC,Yuan2016VTC}, the eavesdropper (Eve) has multiple antennas and the artificial noise (AN) is embed at the terrestrial BS in our design to guarantee the PU. Furthermore, in order to study a more realistic scenario, the case of imperfect CSI is further investigated. Our design objectives are to minimize the transmit power while satisfying secrecy rate constraint at the PU, information rate constraint at the secondary user (SU) and  the total transmit power constraint.
To the best of our knowledge, such work has not been tackled hitherto.
The power minimization problems formulated are non-convex and challenging, which are handled relying on advanced matrix inequality.

\section{Network Model And Problem Formulation\label{b}}

\begin{figure}
\centering
\includegraphics[width=3.0in]{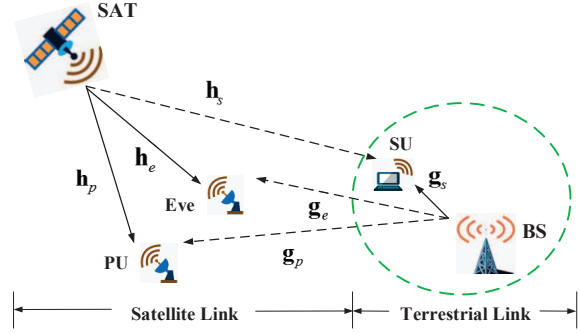}
\caption{The cognitive satellite-terrestrial networks.}
\label{sys}
\end{figure}

The forward link cognitive STN under spectrum sharing is considered in Fig. \ref{sys}, where
the satellite network corresponding to primary network shares the same spectrum with
the terrestrial network corresponding to secondary network. The SAT is equipped with $N_t$ antennas (feeds) and geostationary,
communicating with a satellite terminal (i.e., PU) in the presence of a $N_e$-antenna Eve which just eavesdrops the satellite signal. The terrestrial BS with $N_s$ antennas sends signal to a terrestrial terminal (i.e., SU), while the signal transmitted by the terrestrial BS can be seen as interference to the satellite link. The PU and SU are equipped with single antenna each. Note that the worst-case scenario for eavesdropping under the multi-antenna Eve is taken into account.
The satellite network and terrestrial network are assumed to be operated at the S/L band.

\subsection{Channel Model}
For practical purposes, the satellite links are assumed to experience widely-adopted Shadowed-Rician fading  \cite{An2016JSAC} and the
terrestrial links undergo the correlated Rayleigh fading. Specially, the channels for the satellite links can be respectively modeled as
\begin{align}
\mathbf{h}_i=\sqrt{b(\varphi_i)}\tilde{\mathbf{h}}_i,~\mathbf{H}_e=\sqrt{b(\varphi_e)}\tilde{\mathbf{H}}_e,
\end{align}
with
\begin{align}
b(\varphi)=\left(\frac{J_1(u)}{2u}+36\frac{J_3(u)}{u^3} \right)^2,~u=2.07123\frac{\sin\varphi}{\sin\varphi_{3\mathrm{dB}}},
\end{align}
\begin{align}
\tilde{\mathbf{h}}_i=A\exp(j\boldsymbol{\psi}_i)+Z\exp(j\boldsymbol{\phi}_i),~\tilde{\mathbf{H}}_e=A\exp(j\boldsymbol{\psi}_e)+Z\exp(j\boldsymbol{\phi}_e),
\end{align}
where $b(\varphi)$ is the corresponding beam gain factor, which is determined by their location.
$\varphi$ is the angle between the corresponding receiver and the beam center, and $\varphi_{3\mathrm{dB}}$ is the $3$-dB angle.
$J_1(\cdot)$ and $J_3(\cdot)$ represent the first-kind Bessel function of order 1 and 3.
$\tilde{\mathbf{h}}_i\in \mathbb{C}^{N_t\times 1}~(i=\{p,s\})$ denotes the channel fading vector from SAT to the receiver $i$ and $\tilde{\mathbf{H}}_e\in \mathbb{C}^{N_t\times N_e}$  denotes the channel fading matrix from SAT to Eve, which include the scattering and the line-of-sight (LOS) components.
$\boldsymbol{\psi}\in [0,2\pi)$ denotes the stationary random phase and $\boldsymbol{\phi}$ denotes the deterministic phase of the LOS component.

On the other hand, the terrestrial channel vectors, using the well-known Kronecker model, are modeled as
\begin{align}
\mathbf{g}_i=\mathbf{R}_i^{\frac{1}{2}}\tilde{\mathbf{g}}_i,~\mathbf{G}_e=\mathbf{R}_e^{\frac{1}{2}}\tilde{\mathbf{G}}_e,
\end{align}
where $\tilde{\mathbf{g}}_i\in \mathbb{C}^{N_s\times 1}~(i\in\{p,s\})$ denotes the channel vector from the terrestrial BS to receiver $i$ and  $\tilde{\mathbf{G}}_e\in \mathbb{C}^{N_s\times N_r}$ denotes the channel matrix from the terrestrial BS to Eve, which all follow Rayleigh fading. $\mathbf{R}_i$ and $\mathbf{R}_e$ are the corresponding correlation matrices given by simulation section.
Following \cite{An2016JSAC}, we assume that the terrestrial BS is equipped with uniform linear antenna (ULA) array, and the correlation matrix $\mathbf{R}_i, ~(i = \{p, e, s\})$ with $(m, n)$-th element is given by
\begin{align}
\left[\mathbf{R}_i\right]_{m,n}\approx \frac{1}{2\pi}\int_0^{2\pi} \exp \left[-j2\pi(m-n)\Delta\theta_i\frac{d}{\lambda}\sin\theta_i\right]d\theta
\end{align}
where $\theta_i$ denotes the angle-of departure (AOD) and $\Delta\theta_i$ denotes the angle spread, $d$ is the minimum inter-element spacing and $\lambda$ is the carrier
wavelength.

\subsection{Problem Formulation}
Let $\mathbf{x}_p\in \mathbb{C}^{N_t\times 1}$ and $\mathbf{x}_s\in \mathbb{C}^{N_s\times 1}$ be the transmitted signal vectors at SAT and terrestrial BS, where $\mathbf{x}_p\sim\mathcal{CN} (\mathbf{0},\mathbf{Q}_p)$
is the transmit covariance matrix of $\mathbf{x}_p$.
To provide the stronger distortions to Eve, the AN is adopted at terrestrial BS namely,
$\mathbf{x}_s=\mathbf{s}+\mathbf{z}$,
where $\mathbf{s}\sim\mathcal{CN} (\mathbf{0},\mathbf{Q}_s)$, $\mathbf{Q}_s$ is the transmit covariance matrix of $\mathbf{s}$.
In addition, $\mathbf{z}\sim\mathcal{CN} (\mathbf{0},\mathbf{Q}_z)$ is the AN vector artificially invoked by terrestrial BS. Note that $\mathbf{Q}_z$ represents the transmit covariance matrix of $\mathbf{z}$.
Then, the received signals at PU, Eve and SU are expressed, respectively, as
\begin{align}
  &y_p=\mathbf{h}_p^H\mathbf{x}_p+\mathbf{g}_p^H\mathbf{s}+\mathbf{g}_p^H\mathbf{z}+n_p,
  \label{eq1} \\
  &\mathbf{y}_e=\mathbf{H}_e^H\mathbf{x}_p+\mathbf{G}_e^H\mathbf{s}+\mathbf{G}_e^H\mathbf{z}+\mathbf{n}_e,
  \label{eq2} \\
  &y_s=\mathbf{g}_s^H\mathbf{s}+\mathbf{g}_s^H\mathbf{z}+\mathbf{h}_s^H\mathbf{x}_p+n_s,
  \label{eq3}
\end{align}
where $n_i\sim\mathcal{CN}(0,\sigma_i^2)~(i\in \{p,s\})$ and $\mathbf{n}_e\sim\mathcal{CN}(\mathbf{0}_{N_r},\sigma_e^2\mathbf{I}_{N_r})$ denote additive white Gaussian noises at the receiver $i$ and Eve, respectively.

According to \cite{An2016JSAC,Zheng2012TWC}, the secrecy rate of PU is defined as
\begin{align}
  R_{sec,p}= \Big[C_p(\mathbf{Q}_p,\mathbf{Q}_s,\mathbf{Q}_z)-C_e(\mathbf{Q}_p,\mathbf{Q}_s,\mathbf{Q}_z)\Big]^+,
\end{align}
where $[x]^+=\max\{x,0\}$. $C_p(\mathbf{Q}_p,\mathbf{Q}_s,\mathbf{Q}_z)$ and $C_e(\mathbf{Q}_p,\mathbf{Q}_s,\mathbf{Q}_z)$ denote the information rates at PU and Eve, respectively, i.e.,
\begin{align}
  C_p(\mathbf{Q}_p,\mathbf{Q}_s,\mathbf{Q}_z)= \log_2\left(1+\frac{\mathbf{h}_p^H\mathbf{Q}_p\mathbf{h}_p}{\mathbf{g}_p^H(\mathbf{Q}_s+\mathbf{Q}_z)\mathbf{g}_p+\sigma_p^2}\right),
  \end{align}
\begin{align}
  C_e(\mathbf{Q}_p,\mathbf{Q}_s,\mathbf{Q}_z)= \log_2\det\left(\mathbf{I}+\frac{\mathbf{H}_e^H\mathbf{Q}_p\mathbf{H}_e}{\mathbf{G}_e^H(\mathbf{Q}_s+\mathbf{Q}_z)\mathbf{G}_e+\sigma_e^2\mathbf{I}}\right).
\end{align}

Following the study of \cite{Zhu2016TVT,Yuan2016VTC}, the secondary network can access the spectrum licensed to PU when the PLS of the primary network is guaranteed.
In the considered scenario, the transmit power of the information signal
is minimized subject to the secrecy rate constraint at PU, the information rate constraint at SU and the total transmit power constraint
such that the AN transmit power is maximized for secrecy consideration. As a result, the described problem is formulated as
\begin{subequations}
\begin{align}
&\min_{\mathbf{Q}_p\succeq\mathbf{0},\mathbf{Q}_s\succeq\mathbf{0},\mathbf{Q}_z\succeq\mathbf{0}}~ \mathrm{Tr}(\mathbf{Q}_p)+\mathrm{Tr}(\mathbf{Q}_s)\\
\mbox{s.t.}\quad
   \label{m1}
   &C_p(\mathbf{Q}_p,\mathbf{Q}_s,\mathbf{Q}_z)-C_e(\mathbf{Q}_p,\mathbf{Q}_s,\mathbf{Q}_z) \geq \tau_p,  \\
   \label{m2}
   &\log_2\left(1+\frac{\mathbf{g}_s^H\mathbf{Q}_s\mathbf{g}_s}{\mathbf{g}_s^H\mathbf{Q}_z\mathbf{g}_s+\mathbf{h}_s^H\mathbf{Q}_p\mathbf{h}_s+\sigma_s^2}\right) \geq \tau_s, \\
   \label{m3}
   &\mathrm{Tr}(\mathbf{Q}_p)+\mathrm{Tr}(\mathbf{Q}_s)+\mathrm{Tr}(\mathbf{Q}_z)\leq P_{th},
\end{align}
\label{op1}
\end{subequations}

\noindent where $\tau_p$ and $\tau_s$ represent the prescribed secrecy rate target at PU and information rate target at SU, respectively. $P_{th}$ is the maximum transmit power of the network.
The problem (\ref{op1}) is not convex in terms of its constraints (\ref{m1}) and (\ref{m2}), which cannot be solved efficiently. In order to circumvent this non-convex issue, we propose a joint design of the transmit covariance matrices and AN matrix to relax this problem as the convex optimization framework in the following.

\section{Secure Beamforming Design \label{c}}
In this section, the perfect CSI case is first studied where full CSI of all users
is assumed to be perfectly known at SAT and terrestrial BS. This assumption has been widely adopted in \cite{Zheng2012TWC,Yuan2016VTC}. Then, the research is further extended to the imperfect
CSI case, in which the CSI of PU and Eve is unknown at terrestrial BS since the PU and Eve belong to the satellite network.

\subsection{Perfect CSI\label{c}}
Assuming all CSI is known perfectly at SAT and terrestrial BS,
by introducing a slack variable $\beta$, problem (\ref{op1}) is equivalently rewritten as

\begin{subequations}
\begin{align}
&\min_{\mathbf{Q}_p\succeq\mathbf{0},\mathbf{Q}_s\succeq\mathbf{0},\mathbf{Q}_z\succeq\mathbf{0},\beta} \mathrm{Tr}(\mathbf{Q}_p)+\mathrm{Tr}(\mathbf{Q}_s)\\
\mbox{s.t.}\quad
   \label{q1}
   &\log_2\left(1+\frac{\mathbf{h}_p^H\mathbf{Q}_p\mathbf{h}_p}{\mathbf{g}_p^H(\mathbf{Q}_s+\mathbf{Q}_z)\mathbf{g}_p+\sigma_p^2}\right)-\log_2\beta \geq \tau_p,  \\
   \label{q2}
   &\det\left(\mathbf{I}+\frac{\mathbf{H}_e^H\mathbf{Q}_p\mathbf{H}_e}{\mathbf{G}_e^H(\mathbf{Q}_s+\mathbf{Q}_z)\mathbf{G}_e+\sigma_e^2\mathbf{I}} \right)\leq \beta, \\
   \label{q3}
   &\log_2\left(1+\frac{\mathbf{g}_s^H\mathbf{Q}_s\mathbf{g}_s}{\mathbf{g}_s^H\mathbf{Q}_z\mathbf{g}_s+\mathbf{h}_s^H\mathbf{Q}_p\mathbf{h}_s+\sigma_s^2}\right) \geq \tau_s, \\
      \label{q4}
   &\mathrm{Tr}(\mathbf{Q}_p)+\mathrm{Tr}(\mathbf{Q}_s)+\mathrm{Tr}(\mathbf{Q}_z)\leq P_{th}.
\end{align}
\label{op2}
\end{subequations}
where $\log_2 \beta$ can be interpreted as the maximal allowable mutual information for Eve.
To circumvent the difficulty of constraint (\ref{q2}), we use the following lemma whose proof can be found in \cite{Li2013TSP}.

\textit{Lemma 1}: The following implication holds
\begin{align}
&\det\left(\mathbf{I}+\frac{\mathbf{H}_e^H\mathbf{Q}_p\mathbf{H}_e}{\mathbf{G}_e^H(\mathbf{Q}_s+\mathbf{Q}_z)\mathbf{G}_e+\sigma_e^2\mathbf{I}} \right)\leq \beta \label{perfect1}\\
&\Longrightarrow (\beta-1)\left(\mathbf{G}_e^H(\mathbf{Q}_s+\mathbf{Q}_z)\mathbf{G}_e+\sigma_e^2\mathbf{I}\right)-\mathbf{H}_e^H\mathbf{Q}_p\mathbf{H}_e\succeq \mathbf{0} \label{perfect2}
\end{align}
for any $\mathbf{G}_e\in \mathbb{C}^{N_s\times N_r}$, $\mathbf{H}_e\in \mathbb{C}^{N_t\times N_r}$, $\mathbf{Q}_p\succeq\mathbf{0}$, $\mathbf{Q}_s\succeq\mathbf{0}$ and $\mathbf{Q}_e\succeq\mathbf{0}$.
Furthermore, the equivalence in (\ref{perfect1}) and (\ref{perfect2}) hold if $\mathrm{rank}(\mathbf{Q}_p)\leq 1$ and $\mathrm{rank}(\mathbf{Q}_s)\leq 1$.

Replacing (\ref{perfect1}) with (\ref{perfect2}), the optimization problem given by (\ref{op1}) is reformulated as
\begin{subequations}
\begin{align}
&\min_{\mathbf{Q}_p\succeq\mathbf{0},\mathbf{Q}_s\succeq\mathbf{0},\mathbf{Q}_z\succeq\mathbf{0},\beta} \mathrm{Tr}(\mathbf{Q}_p)+\mathrm{Tr}(\mathbf{Q}_s)\\
\mbox{s.t.}\quad
   \label{p1}
   &\mathrm{Tr}\left(\alpha \mathbf{G}_p(\mathbf{Q}_s+\mathbf{Q}_z)\right)+\mathrm{Tr}(\mathbf{H}_p\mathbf{Q}_p)+\alpha \sigma_p^2 \geq 0,  \\
   \label{p2}
   &(\beta-1)\left(\mathbf{G}_e^H(\mathbf{Q}_s+\mathbf{Q}_z)\mathbf{G}_e+\sigma_e^2\mathbf{I}\right)-\mathbf{H}_e^H\mathbf{Q}_p\mathbf{H}_e\succeq \mathbf{0}, \\
   \label{p3}
   &\mathrm{Tr}\left(\mathbf{G}_s(\mathbf{Q}_s-\gamma\mathbf{Q}_z)\right)-\gamma\left(\mathrm{Tr}(\mathbf{H}_s\mathbf{Q}_p)+\sigma_s^2\right)\geq 0, \\
      \label{p4}
   &\mathrm{Tr}(\mathbf{Q}_p)+\mathrm{Tr}(\mathbf{Q}_s)+\mathrm{Tr}(\mathbf{Q}_z)\leq P_{th},
\end{align}
\label{op3}
\end{subequations}
where $\alpha=1-\beta2^{\tau_p}$ and $\gamma=2^{\tau_s}-1$, meanwhile $\mathbf{H}_p=\mathbf{h}_p\mathbf{h}_p^H$, $\mathbf{G}_p=\mathbf{h}_p\mathbf{h}_p^H$,
$\mathbf{H}_s=\mathbf{h}_s\mathbf{h}_s^H$ and $\mathbf{G}_s=\mathbf{g}_s\mathbf{g}_s^H$.
Notably, we observe that the problem given by (\ref{op3}) is a convex SDP problem when $\beta$ is fixed, which can be efficiently solved by the interior point method \cite{Boyd2004}. Then, the one-dimensional linear
search is performed to find the optimal $\beta^*$ from the interval $[1,1+P_{th}||\mathbf{h}_p||^2]$.

\subsection{Extension to Imperfect CSI\label{c}}

Since PU and Eve are the users of the satellite network, it is very difficult
for the terrestrial BS to obtain accurate CSI of PU and Eve. Thus, the CSI imperfectness between BS-PU and BS-Eve links has to be taken into
consideration in the network design.
In this setup, the imperfect CSI from BS to PU and Eve are modeled, respectively, as
\begin{align}
\mathbf{g}_p=\hat{\mathbf{g}}_p+\Delta\mathbf{g}_p, ~
\mathbf{G}_e=\hat{\mathbf{G}}_e+\Delta\mathbf{G}_e,\label{robust1}
\end{align}
where $\hat{\mathbf{g}}_p$ and $\hat{\mathbf{g}}_e$ are the nominal channel vectors. And $\Delta\mathbf{g}_p$ and $\Delta\mathbf{g}_e$ denote the channel estimation errors, which are
respectively bounded by
\begin{align}
\|\Delta\mathbf{g}_p\|_2=\|\mathbf{g}_p-\hat{\mathbf{g}}_p\|_2\leq\epsilon_p,
\|\Delta\mathbf{G}_e\|_F=\|\mathbf{G}_e-\hat{\mathbf{G}}_e\|_F\leq\epsilon_e,\label{robust2}
\end{align}
where $\epsilon_p$ and $\epsilon_e$ represent the non-negative channel error bounds.

Based on (\ref{robust1}) and (\ref{robust2}), the robust secure transmission for minimizing total transmit power is designed. Mathematically, the joint optimization problem is expressed as
\begin{subequations}
\begin{align}
&\min_{\mathbf{Q}_p\succeq\mathbf{0},\mathbf{Q}_s\succeq\mathbf{0},\mathbf{Q}_z\succeq\mathbf{0},\bar{\beta}} \mathrm{Tr}(\mathbf{Q}_p)+\mathrm{Tr}(\mathbf{Q}_s)\\
\mbox{s.t.}
   \label{x1}
   &\min_{\|\Delta\mathbf{g}_p\|_2\leq\epsilon_p}\frac{\mathbf{h}_p^H\mathbf{Q}_p\mathbf{h}_p+\mathbf{g}_p^H(\mathbf{Q}_s+\mathbf{Q}_z)\mathbf{g}_p+\sigma_p^2}{\bar{\beta}\left(\mathbf{g}_p^H(\mathbf{Q}_s+\mathbf{Q}_z)\mathbf{g}_p+\sigma_p^2\right)} \geq 2^{\tau_p},  \\
   \label{x2}
   &\max_{\|\Delta\mathbf{G}_e\|_F\leq\epsilon_e}\det\left(\mathbf{I}+\frac{\mathbf{H}_e^H\mathbf{Q}_p\mathbf{H}_e}{\mathbf{G}_e^H(\mathbf{Q}_s+\mathbf{Q}_z)\mathbf{G}_e+\sigma_e^2\mathbf{I}} \right)\leq \bar{\beta}, \\
   \label{x3}
   &\mathrm{Tr}\left(\mathbf{G}_s(\mathbf{Q}_s-\gamma\mathbf{Q}_z)\right)-\gamma\left(\mathrm{Tr}(\mathbf{H}_s\mathbf{Q}_p)+\sigma_s^2\right)\geq 0, \\
      \label{x4}
   &\mathrm{Tr}(\mathbf{Q}_p)+\mathrm{Tr}(\mathbf{Q}_s)+\mathrm{Tr}(\mathbf{Q}_z)\leq P_{th},
\end{align}
\label{robustEq1}
\end{subequations}
where $\bar{\beta}$ is the introduced variable similar to the perfect case.
Note that problem (\ref{robustEq1}) is costly to solve globally since
there infinitely many constraints in (\ref{x1}) and (\ref{x2}). To make problem (\ref{robustEq1}) computationally tractable, we convert these
robust constraints into linear matrix inequalities (LMIs) relying on advanced matrix inequality results.

To proceed, applying $\mathcal{S}$-\textit{Procedure} \cite{Boyd2004} and introducing a slack variable $\mu_p\geq 0$, we convert (\ref{x1}) into an LMI given by (\ref{robustlongEq1})
\begin{small}
\begin{equation}
\mathbf{\Gamma}_p(\mathbf{Q}_p,\mathbf{Q}_s,\mathbf{Q}_z,\bar{\beta},\mu_p)=
\begin{bmatrix}
\mu_p\mathbf{I}_{N_s}+\bar{\alpha}(\mathbf{Q}_s+\mathbf{Q}_z) &\bar{\alpha}(\mathbf{Q}_s+\mathbf{Q}_z)\hat{\mathbf{g}}_p  \\
\hat{\mathbf{g}}_p^H\bar{\alpha}(\mathbf{Q}_s+\mathbf{Q}_z) & \hat{\mathbf{g}}_p^H\bar{\alpha}(\mathbf{Q}_s+\mathbf{Q}_z)\hat{\mathbf{g}}_p+\mathbf{h}_p^H\mathbf{Q}_p\mathbf{h}_p+\bar{\alpha}\sigma_p^2-\mu_p\epsilon_p^2
\end{bmatrix}
\label{robustlongEq1}
\end{equation}
\end{small}
where $\bar{\alpha}=1-\bar{\beta}2^{\tau_p}$.
Then, to circumvent this difficulty of constraint (\ref{x2}), we introduce the following lemma, i.e.,
the extension of $\mathcal{S}$-\textit{Procedure} \cite{Luo2004SIAM}.

\textit{Lemma 2 (Luo-Sturm-Zhang \cite{Luo2004SIAM}):} Let $f(\mathbf{X})=\mathbf{X}^H\mathbf{A}\mathbf{X} + \mathbf{X}^H\mathbf{B}+\mathbf{B}^H\mathbf{X}+\mathbf{C}$ and
$\mathbf{D}\succeq\mathbf{0}$, the following equivalence holds if there exists a $\mu_e \geq 0$:
\begin{align}
&f(\mathbf{X})\succeq \mathbf{0},~\forall \mathbf{X}\in \{\mathbf{X}|\mathrm{Tr}(\mathbf{D}\mathbf{X}\mathbf{X}^H)\leq 1 \} \nonumber \\
&\Longleftrightarrow
\begin{bmatrix}
\mathbf{C} & \mathbf{B}^H \\
\mathbf{B} & \mathbf{A}
\end{bmatrix}
-\mu_e
\begin{bmatrix}
\mathbf{I}_2 & \mathbf{0} \\
\mathbf{0}_2 & \mathbf{D}
\end{bmatrix}
\succeq \mathbf{0}
\end{align}

Based on Lemma $3$, we have the following proposition.

\textit{Proposition 1:} The following implication holds
\begin{align}
&\max_{\|\Delta\mathbf{G}_e\|_F\leq\epsilon_e}\det\left(\mathbf{I}+\frac{\mathbf{H}_e^H\mathbf{Q}_p\mathbf{H}_e}{\mathbf{G}_e^H(\mathbf{Q}_s+\mathbf{Q}_z)\mathbf{G}_e+\sigma_e^2\mathbf{I}} \right)\leq \bar{\beta} \label{robustEq31}\\
&\Longrightarrow \mathbf{\Gamma}_e(\mathbf{Q}_p,\mathbf{Q}_s,\mathbf{Q}_z,\bar{\beta},\mu_e)\succeq \mathbf{0} \label{robustEq4}
\end{align}
where $\mathbf{\Gamma}_e(\mathbf{Q}_p,\mathbf{Q}_s,\mathbf{Q}_z,\bar{\beta},\mu_e)$ is given by
\begin{footnotesize}
\begin{equation}
\mathbf{\Gamma}_e(\mathbf{Q}_p,\mathbf{Q}_s,\mathbf{Q}_z,\bar{\beta},\mu_e)=
\begin{bmatrix}
(\bar{\beta}-1-\mu_e)\mathbf{I}_{N_r}+\hat{\mathbf{G}}_e^H\left[(\bar{\beta}-1)(\mathbf{Q}_s+\mathbf{Q}_z)\right]\hat{\mathbf{G}}_e-\mathbf{H}_e^H\mathbf{Q}_p\mathbf{H}_e &\hat{\mathbf{G}}_e^H(\beta-1)(\mathbf{Q}_s+\mathbf{Q}_z)  \\
(\bar{\beta}-1)(\mathbf{Q}_s+\mathbf{Q}_z)\hat{\mathbf{G}}_e & (\bar{\beta}-1)(\mathbf{Q}_s+\mathbf{Q}_z)+\frac{\mu_e}{\epsilon_p^2}\mathbf{I}_{N_s}
\end{bmatrix}
\label{robustlongEq2}
\end{equation}
\end{footnotesize}
Furthermore, the equivalence in (\ref{robustEq31}) and (\ref{robustEq4}) hold if $\mathrm{rank}(\mathbf{Q}_p)\leq 1$ and $\mathrm{rank}(\mathbf{Q}_z)\leq 1$.
Due to the space limitation, we omit the proof, which can be referred to \cite[Proposition 2] {Li2013TSP} for more details.

Replacing (\ref{x1}) and (\ref{x2}) with (\ref{robustlongEq1}) and (\ref{robustlongEq2}), problem (\ref{robustEq1}) can be relaxed to the following problem as
\begin{subequations}
\begin{align}
\min_{\mathbf{Q}_p,\mathbf{Q}_s,\mathbf{Q}_z,\bar{\beta},\mu_p,\mu_e}&~\mathrm{Tr}(\mathbf{Q}_p)+\mathrm{Tr}(\mathbf{Q}_s)  \\
\mbox{s.t.}\quad
   &(\mathrm{\ref{x3}}), ~(\mathrm{\ref{x4}}), ~(\ref{robustlongEq1}),~(\ref{robustlongEq2}), \\
   &\mathbf{Q}_p\succeq\mathbf{0},~\mathbf{Q}_s\succeq\mathbf{0},~\mathbf{Q}_z\succeq\mathbf{0}.
\end{align}
\label{robustEq3}
\end{subequations}
Noticeably, problem (\ref{robustEq3}) is a convex SDP when $\bar{\beta}$ is fixed, which can be efficiently solved by a standard optimization solver, e.g. CVX \cite{Boyd2004}. Then, the one-dimensional linear search is performed to find the optimal $\beta^*$ from the interval $[1,1+P_{th}||\mathbf{h}_p||^2]$.

\textit{Remark $1$:} For the two cases, if the optimal solutions $\mathbf{Q}_p^*$ and $\mathbf{Q}_s^*$ are rank-one, the optimal beamforming vectors can be obtained via eigenvalue decomposition.
Surprisingly, as observed from the simulations with $1000$ randomly generated channel realizations, the obtained optimal solutions $\mathbf{Q}_p^*$ and $\mathbf{Q}_s^*$ for the relaxed (\ref{op3}) are always rank-one. Also, the obtained rank-one optimal solutions $\mathbf{Q}_p^*$ and $\mathbf{Q}_s^*$ for the relaxed (\ref{robustEq3}) are about $99\%$ and $91.4\%$, respectively.
Alternatively, Gaussian randomization is applied to obtain the approximate solutions.

\textit{Complexity Analysis:} According to \cite{Boyd2004}, the main computational complexity for solving problem (\ref{op3}) is $\mathcal{O}\left(3^4(N_s^{1/2}+N_t^{1/2})\right)\log_2\left(\frac{1}{\varepsilon}\right)$,
and for solving problem (\ref{robustEq3}) is $\mathcal{O}\left((N_s+1)^{1/2}+(N_s+N_r)^{1/2}+(N_s^{1/2}+N_t^{1/2})\right)\\ \log_2\left(\frac{1}{\varepsilon}\right)$, respectively, where $\varepsilon$ is the given accuracy.

\section{Numerical Results \label{e}}

This section evaluates the performance of the proposed schemes.
The SAT is assumed to be equipped with $N_t = 4$ antennas (feeds) to cover the network. The number of antennas equipped at terrestrial BS and Eve are $N_{s}=4$ and $N_{r}=2$, respectively.
The satellite links experience heavy shadowing with parameters $(b_i, m_i, \Omega_i)= (0.063, 2, 8.97 \times 10^{-4}), \forall i\in\{p,e,s\}$.
The beam angles from SAT to PU, to Eve and to SU are set as $0.01^\circ$, $0.8^\circ$ and $30^\circ$, respectively.
For simplicity, we assume that the noise powers are identical as $\sigma_p^2=\sigma_e^2=\sigma_s^2=1$ and we set $\epsilon_p=\epsilon_e=\epsilon$ and $P_{th}=60$W,
$d = \frac{\lambda}{2}$, $ \theta_s= 0^\circ$, $\theta_p = 40^\circ$, and $\Delta\theta_s= \Delta\theta_e =5^\circ$.

\begin{figure}
\centering
\includegraphics[width=2.7in]{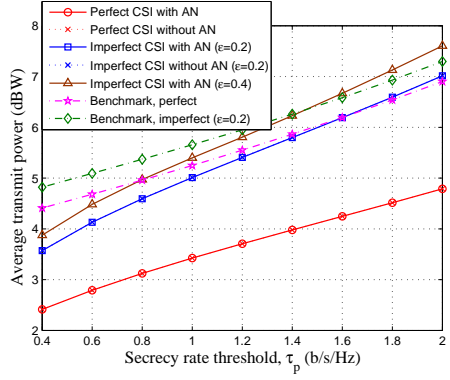}
\caption{Average transmit power versus secrecy rate threshold at PU with $2^{\tau_s}=10$.}
\label{fig:PF1}
\end{figure}

Fig. \ref{fig:PF1} shows the total transmit power comparison of the proposed cooperative beamforming scheme, without AN scheme and a fixed allocation scheme (as a benchmark).
As expected, we can clearly see that the proposed design schemes with AN outperform the benchmark schemes for the same secrecy rate threshold $\tau_p$  saving  over
$2$ dBW in the secrecy regions due to the optimized transmission. Furthermore, we can observe that a higher power
consumption is required for the imperfect CSI than that under the perfect CSI case.
As channel error $\epsilon$ increases, the performance loss becomes larger, which implies that the robust design scheme is very sensitive to the CSI accuracy.
In addition, it can be interestingly observed that the proposed optimal schemes with and without
AN are almost the same. This is due to the fact that the main advantage of using AN can be utilized to reduce the number of required
antenna at terrestrial BS when existing at least two eavesdroppers (or visualized as a multi-antenna Eve), according to the Theorem $2$ in \cite{Zheng2012TWC}.
\begin{figure}
\centering
\includegraphics[width=2.8in]{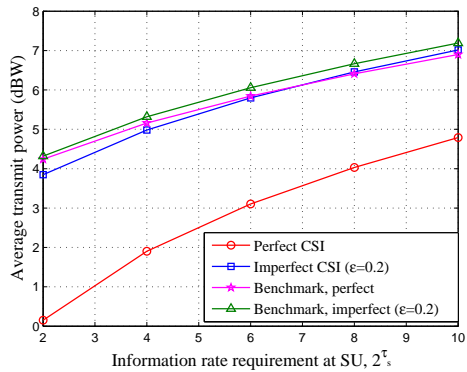}
\caption{Average transmit power versus information rate requirement at SU.}
\label{fig:PF2}
\end{figure}

Fig. \ref{fig:PF2} plots the total transmit power versus information rate requirement at SU under $\tau_p=2$ b/s/Hz.
The curves in Fig. \ref{fig:PF2} demonstrate that the total transmit power increases with the
increasing of $2^{\tau_s}$. This is mainly due to the fact that, with increasing $2^{\tau_s}$, more power will be used to satisfy the minimum information rate
requirement for SU.

\section{Conclusions \label{f}}

In this letter, we investigated the secure cooperative transmission scheme for the cognitive STNs with the objective of minimizing the transmit power of information signal, while satisfying the required constraints.
Both the perfect and imperfect CSI were considered in the design.
Simulation results showed that the proposed schemes have a good performance in guaranteeing secure transmission.

\bibliographystyle{IEEEtran}
\bibliography{refs}

\end{document}